\documentclass[showpacs,showkeys,preprintnumbers,preprint,amsmath,amssymb,superscriptaddress]{revtex4}

\usepackage{graphicx}
\usepackage{dcolumn}
\usepackage{bm}
\begin{document}

\title{The maximum entropy production principle and linear irreversible processes}

\author{Pa\v{s}ko \v{Z}upanovi\'{c}}
 \email{pasko@pmfst.hr}
\author{Domagoj Kui\'{c}}
 \email{dkuic@pmfst.hr}
 \author{ \v{Z}eljana Bona\v{c}i\'{c} Lo\v{s}i\'{c} }
 \email{agicz@pmfst.hr}
 \author{ Dra\v{z}en Petrov }
 \email{drapet@pmfst.hr}
 \author{Davor Jureti\'{c}}
 \email{juretic@pmfst.hr}
\affiliation
{ University of Split, Faculty of Science, Teslina 12,  21000 Split,  Croatia }
\author{Milan Brumen}
\email{Milan.Brumen@uni-mb.si }
\affiliation{University of Maribor, Faculty of Natural Sciences and Mathematics\\ Koro\v{s}ka cesta 160, SI-2000 Maribor, Slovenia\\
 Jo\v{z}ef Stefan Institute \\ Jamova cesta 39, SI-1000 Ljubljana, Slovenia } 

\date{\today}




\begin{abstract}
It is shown  that Onsager's principle of the least dissipation of energy
   is equivalent to the maximum entropy production principle.
   It is known that  solutions of the linearized Boltzmann equation make extrema of   entropy production. 
   It is argued, in the case of stationary processes,  that this extremum  is a maximum rather than  a minimum.  
\end{abstract}
\maketitle

\section{Introduction}

It was Ehrenfest (Enzykl. Math. Wissensch, IV, 2(II) fasc.6, p82, note23, 1912)
who first asked whether a function exists which, like  entropy
in the equilibrium state of an isolated system, achieves its extreme value in a
stationary non-equilibrium state.

There are various results and formulations  of irreversible thermodynamics based on the extremum of entropy production. They are related either to the minimum or maximum entropy production.
The well known  example based on minimum entropy production is Prigogine theorem  \cite{prigogine}.

There are several different results that concern maximum entropy production (MEP).  Ziegler applied the  principle of maximum  entropy production  in thermomechanics \cite{zieglerbook}. It states that the rate of entropy production under prescribed forces should be maximum. Paltridge \cite{paltridge} has formulated empirically  MEP principle in order to describe the Earth climate. Apart from details of his model  (like  albedo or  cloudiness) the main features of his model are steady incoming sun radiation, outgoing radiation proportional to the fourth power of the local annual average temperature and horizontal heat flux from equator to poles.  There are no boundary conditions on atmosphere. Paltridge has proposed that steady state is the state of maximum entropy production due to the  latitudinal heat transport. His  predictions of  the annually average distribution of the temperature of atmosphere ( and some other climatic parameters)   fit very well measured values.  Kohler \cite{kohlerM} has started from the Boltzmann's transport equation and has  shown that the solution for the velocity distribution function  of rarefied gas, in stationary state close to the equilibrium,  is the state of extremum entropy production. He has  argued that   extremum type  depends on the choice of the constraint. We show in this paper that constraint that leads to the stationary state as the state of the minimum entropy production is not consistent with his starting assumption. We argue here that   rarefied gas in the stationary state close to equilibrium is in the MEP state. 

Recently Jaynes' principle of maximum information entropy (MaxEnt) has been exploited to derive the MEP principle. Dewar \cite{dewar1,dewar2} has introduced paths as possible  trajectories of the system in    phase space. Then  using the  MaxEnt procedure he has shown , under some assumptions,  that the most probable development of the system is accompanied with MEP. Niven \cite{niven} has applied MaxEnt using the values of fluxes as variables  and he  found that the stationary state of the system is  the MEP state. 

At first sight minimum and maximum entropy production results seem to contradict each other. But if one considers   starting assumptions one finds that these assumptions are  different. Thus these results are independent of one another \cite{ziman,martjushev}

In this paper we focus on Onsager's principle of least dissipation of energy and show that this principle is equivalent to MEP principle. Secondly, as we have already noted, we show that Kohler interpretation of stationary state of rarefied gas close to the equilibrium should be restricted only on the constraint that it is consistent with the interpretation of this state as the  MEP state.

This is the first of three papers in a series and are referred to as papers I, II and III. In paper II we discuss whether stationary or relaxation processes are suitable for the formulation of 
 principles. We argue in favour of relaxation processes. In paper III we apply MaxEnt formalism to relaxation processes and we derive the MEP principle as its corollary.

\section{The principle of the least dissipation of energy and linear nonequilibrium thermodynamics}

Onsager's famous   papers \cite{onsager1,onsager2} examine   linear nonequilibrium thermodynamics. The first sections of these papers examine reciprocal relations. 
The following sections are devoted to the formulation of linear nonequilibrium thermodynamics. Onsager has  employed the principle of the least dissipation of energy.  
 In the second part of  his second paper \cite{onsager2}, Onsager has applied the principle of the least dissipation energy to the general linear nonequilibrium process. 
 
  A linear   relationship between $n$ independent thermodynamic forces   $\{X_i\}$ and their  conjugated fluxes $\{j_i\}$,
\begin{equation}
\label{system}
	X_i=\sum_j  R_{ij} j_j,
\end{equation}
 exists for  the system  close to equilibrium.
  The density of the entropy production is equal to the product of the thermodynamic forces and conjugated fluxes \cite{prigogine}, \cite{degroot}. The total entropy production is
\begin{equation}
\label{EP}
\sigma =\int \sum^n_{j=1} X_j j_j dV.
\end{equation}
 In his second famous paper Onsager \cite{onsager2} has introduced the dissipation function 
\begin{equation}
\label{dis}
	\Phi= \frac{1}{2} \int \sum^n_{i,j=1}  R_{ij}  j_i j_j dV,
\end{equation}
and formulated 
the principle of the least dissipation of energy,
\begin{equation}
	\sigma-	\Phi= maximum.
\end{equation}

The variational procedure
\begin{equation}
\delta \left[\sigma-	\Phi\right]= 0,
\end{equation}
 gives
  the system of equations (\ref{system}).
 In other words,  the linear irreversible thermodynamics can be inferred  from  the principle of the least dissipation of energy.

 \section{The MEP principle  and linear nonequilibrium thermodynamics}

An alternative approach to linear nonequilibrium thermodynamics is based on the phenomenological fact that nonequilibrium processes are characterized by fluxes. 
Therefore, physical quantities
 relevant for a description of the time development of the system must be functions of fluxes. The standard approach to the nonequilibrium thermodynamics is based on the laws of conservation of mechanical physical quantities mass, momentum and energy. The next step is to add heat as the additional mechanism of the  exchange of energy between systems. Then assuming local equilibrium one comes  to the density of  entropy production written as the sum of products of heat flux  and thermal thermodynamic force and viscous pressure tensor and strain tensor divided by temperature. This result serves as the basis for the canonical form of the  second postulate of irreversible thermodynamics that reads: Entropy production can be always written as the sum of products of thermodynamic forces and corresponding fluxes \cite{evansbook,degroot}. Thermodynamic forces and fluxes are not independent quantities. In principle fluxes determine forces and vice versa.
Having in mind  this fact and the second postulate of irreversible thermodynamics 
 we can say that  the dynamical state of the system is described only   by  fluxes, or equivalently by thermodynamic forces.

Entropy production is a basic,   characteristic quantity   of a nonequilibrium state. 
If the system is close to the  equilibrium  state we can make the Taylor expansion of the  density of entropy production	$d\sigma/dV$
  in fluxes up to the second order,
\begin{equation}
\label{taylor}
\frac{d \sigma}{dV}=A+ \sum_{i}B_i j_i+   \sum_{i,j}R_{ij} j_ij_j.
\end{equation}
 Here  $j_i$ is the mean value of the $i^{th}$ flux. 
  The first term on the right-hand side of this equation  vanishes since there is no entropy production in
  the  equilibrium state. Coefficients $B_i$ and $R_{ij}$ are the property of the system in equilibrium state.  Entropy production does not depend on the  direction   of the flux flow,
   that is to say,  it must be invariant under the replacement of  $\{j_i \} \rightarrow \{-j_i \}$. This 
   means that coefficients that  multiply  the odd power of fluxes vanish, that is,  $B_i=0$. In the 
   lowest order,  entropy production is the bilinear function of fluxes,
\begin{equation}
\label{square}
\sigma=\sum_{i,j} \int_V  R_{ij} j_ij_j d{\bf r} >0.
\end{equation}

Comparison of Eqs. (\ref{square}) and (\ref{dis}) shows that  the dissipation function  is in fact entropy production written in the space of fluxes.
When a  system is close to equilibrium with locally well defined intensive thermodynamic quantities 
the entropy production is given by Eq.  (\ref{EP}),
 \cite{prigogine}, \cite{degroot}.  
 
It is  pointed in the introductory part of this Section that  canonical form of entropy production (\ref{EP}) comes from the law of conservation of energy. Then  equation,  
\begin{equation}
\label{constraint}	 
\int_V \left(  \sum_i   X_i j_i -    \sum_{i,j}  R_{ij} j_ij_j  \right)d{\bf r} =0.
\end{equation}
is the constraint imposed the entropy production due to the law of conservation of energy.

 We seek the maximum entropy production  (\ref{square}) taking into account the constraint 
 (\ref{constraint}). The standard procedure \cite{krasnov}  is to find the maximum of the functional 
 \begin{equation}
\label{F}	
  F= \sum_{i,j} \int_V  R_{ij} j_ij_j d{\bf r} + \lambda \int_V \left[   
  \sum_i   X_i j_i -   \sum_{i,j}  R_{ij} j_ij_j  \right]d{\bf r}.
\end{equation}
A standard variational calculus of extremum values    combined with the  constraint (\ref{constraint}) gives $\lambda =2$ and $F_l$ becomes
 \[ F= -\sum_{i,j} \int_V  R_{ij} j_ij_j d{\bf r} + 2 \int_V   
  \sum_i   X_i j_i d{\bf r} . \]
This is an equation of a quadratic surface  turned upside down. 
The corresponding extremum of $F$ is maximum. The system of equations that determines the point of maximum ,  $\partial F/ \partial  j_i=0$, is just a system 
of linearly coupled thermodynamic forces and 
 fluxes (\ref{system}).

 From Eq. (\ref{square}) follows $R_{ij}=R_{ji}$ i.e. Onsager's reciprocal relations.
In short,
  linear 
 nonequilibrium thermodynamics and  Onsager's reciprocity relations follow from the MEP principle.

A problem analogous to this is the problem of biochemical cycle kinetics close to the equilibrium state.   Starting from the assumption that 
 the energy conservation law is valid for a whole  network of biochemical reactions we have showed that fluxes are distributed in such a way to produce maximum entropy
  \cite{ZJ}.

 In references  \cite{zbjpr,bzj} we have considered a  linear planar electric network held at a constant temperature. There is only a temperature gradient between network and surroundings. Physically this can be achieved using thermally high and poor  conducting material for a network and surroundings, respectively. There is no coupling between electric and heat currents. Only electric currents are coupled via  electromotive forces. Assuming that the stationary state is the state of maximum possible generated  heat   the Kirchhoff loop law  is derived \cite{zbjpr,bzj}. The principle of the maximum heat dissipation is closely related to the MEP principles. 
  
Using mesh currents   the law of charge conservation  has been taken implicitly. The  energy conservation law is used explicitly  as  the constraint. If one only  takes charge conservation law one comes to the conclusion that stationary state is state of minimum generated heat \cite{jaynes4}. We shell dwell more on this question in the  paper II submitted to this special issue.

 We note that two of us   (P.\v{Z}. and D.J.)   have considered the heat flow in the anisotropic crystal. It is  shown in this  special example that the principle of the least dissipation of energy 
 applied  by Onsager in reference \cite{onsager1} is equivalent to the MEP principle \cite{fizika}.

\section{The linearized Boltzmann equation and the extremum of entropy production}

The important influence of constraints at extrema of entropy production can be
seen in the case of linearized Boltzmann equation.

Here we follow the elegant approach given in reference \cite{martjushev}. 
The Boltzmann equation is valid for rarefied gas where collisions are very well defined events. One seeks the one-particle distribution function $f(\textbf{r},\textbf{v},t)$ satisfying the equation, 
\begin{equation}
\label{be}
	\frac{d f}{d t}= \frac{\partial f}{\partial t}+\textbf{v}\frac{\partial f}{\partial \textbf{\textbf{r}}}+\frac{\textbf{F}}{m}\frac{\partial f}{\partial \textbf{\textbf{v}}}=I.
\end{equation}
Here  the two terms on the left-hand side of equation describe the change in the number of molecules in a given element of the phase space due to the collisionless motion of the molecule in the outer field $\textbf{F}$.  The right-hand side of the equation describes the net change in the number of molecules in a given element of the velocity space due to molecule collisions.
 Assuming the instantaneous change of molecule velocities in the collisions   and 
  taking into account the conservation laws in collisions the integral can be written in the form  \cite{rivkin}
\begin{equation}
	\label{insud}
I= \int \int  (\tilde{f} \tilde{f}'-ff')q d\sigma(q,\textbf{e})d\textbf{v}'.
\end{equation}
Here  $  f=f(\textbf{r},\textbf{v},t) \;; f'=f(\textbf{r},\textbf{v'},t)$
 and 
$\tilde{f} =f(\textbf{r},\tilde{\textbf{v}},t) \;; \tilde{f}' =f(\textbf{r},\tilde{\textbf{v}}',t) $
 are the distribution functions of particles before and after collision, respectively. In addition, $ q=v'-v$    is the magnitude of  relative velocity of particles before collisions,
  $\textbf{e}=(\tilde{\textbf{v}}-\textbf{v})/|\tilde{v}-v|$      is a unit vector parallel to the velocity change of one  particle in collision
 and $d\sigma(q,\textbf{e})$ is the differential cross section. It is known from collision theory that  differential cross section depends on interaction potential,  relative velocity $q$  and unit vector \textbf{e}, i.e. on the scattering angle  \cite{rivkin}.

In the case of local equilibrium, intensive variables (temperature, concentration) are  well defined functions of the space. Then one assumes the approximate solution of Eq. (\ref{be})
\cite{kohlerM}, \cite{martjushev}.
\begin{equation}
	\label{les}
f=f_{0}(1+\Psi(\textbf{v})),
\end{equation}
where
\begin{equation}
f_{0}=n(\frac{m}{2\pi kT})^{3/2}exp(-\frac{mV^{2}}{2kT})
	\label{mb}
\end{equation}
is the Maxwell-Boltzmann distribution.

Due to the assumption of local equilibrium the perturbed function must not  contribute to the prescribed intensive parameters like temperature, mean velocity  and density. Then 
\begin{equation}
	\label{uvjetiza les}
\int f_{0}\Psi(\textbf{v})d\textbf{v}=0, \;\; \int f_{0}\Psi(\textbf{v})Vd\textbf{v}=0, \;\;  \int f_{0}\Psi(\textbf{v})\frac{mV^{2}}{2}d\textbf{v}=0.
\end{equation}

Now the collision integral up to the first order of the perturbed  term becomes
\begin{equation}
	\label{insud2}
I(f)= \int \int \int  f_{0} f'_{0}(\tilde{\Psi}+\tilde{\Psi}'-\Psi-\Psi')qd\sigma(q,\textbf{e})d\textbf{v}'=\hat{O}\Psi.
\end{equation}
The designation of $\Psi$ functions is the same as the distribution functions in Eq. (\ref{insud}).

Operator $\hat{O}$ \cite{kohlerM}, \cite{martjushev} is linear and has the following properties
\begin{equation}
	\label{svojstvo_O1}
\hat{O}( \alpha A+\beta B)=\alpha \hat{O}A+\beta \hat{O}B,
\end{equation}

\begin{equation}
	\label{svojstvo_O2}
\int A \hat{O}Bd\textbf{v}=\int B \hat{O}Ad\textbf{v},
\end{equation}

\begin{equation}
	\label{svojstvo_O3}
\int A \hat{O}Ad\textbf{v}\geq0.
\end{equation}

If we designate the left-hand side of Boltzmann equation (\ref{be}) with $-Z$ it becomes
\begin{equation}
	\label{be1}
Z=-\hat{O} \Psi.
\end{equation}

Multiplying the linearized Boltzmann equation with $-k \ln f$ and integrating over velocity space we get 
\begin{equation}
\label{entropycons}
\frac{\partial s}{\partial t} + \nabla \cdot \textbf{j}_s=k \int \Psi\hat{O}\Psi d\textbf{v}.
\end{equation}

Here $s= -k \int f \ln f d\textbf{v}$ is the density of entropy and $\textbf{j}_s=-k \int\textbf{v} f \ln f d\textbf{v}$ is the flux of entropy. 
Entropy  looks like a fluid.  The left-hand side of Eq. (\ref{entropycons}) is the  total time change of the density of entropy, i.e. it is the density of the entropy production. Then, according to the aforementioned  equation  collisions between molecules described by  function $\Psi$ are the sources of entropy production. 
There are a lot of   functions  obeying conditions (\ref{uvjetiza les}). 
We choose those   distribution functions  that obey the additional condition 
\begin{equation}
	\label{dokaz1}
 \int YZ d\textbf{v}=-\int Y \hat{O} Yd\textbf{v}.
\end{equation}
Function $Y$ is not the solution of the Boltzmann equation, i.e. 
\begin{equation}
	\label{be1}
Z \neq -\hat{O} Y.
\end{equation}
$Y$ can be interpreted as a trial distribution function. The corresponding entropy production due to this distribution is
\begin{equation}
\label{entropyconsY}
\frac{\partial s_{Y}}{\partial t} + \nabla \cdot (\textbf{j}_s)_Y=k [Y,\hat{O}Y].
\end{equation}
If the temperature gradient in $x$ direction is the only thermodynamic force, the left-hand side of  the  equation (\ref{dokaz1}) can be written as \cite{kohlerM}
\begin{equation}
	\label{entropy}
\int YZd\textbf{v}=	-\frac{ (q_x)_Y}{T^2}\frac{\partial T}{\partial x}.
\end{equation}
Here $(q_x)_Y$ is the heat flux associated  with the distribution function $Y$.
This equation gives a physical meaning to condition (\ref{dokaz1}). Varied distribution functions are selected in such a way that the entropy production due to the molecular collisions is equal to the entropy production due to the heat conduction. In other words entropy produced by molecular collisions is equal to the product of the heat flux and conjugated thermodynamic force. 
This condition is in accordance with the starting assumption of a stationary  process.

The multiplication of  the Boltzmann equation with the perturbative distribution function $\Psi$ and integration over molecular velocity space \textbf{v}
 gives
\begin{equation}
	\label{prosirenaBE1}
\int \Psi Zd\textbf{v}=-\int \Psi \hat{O} \Psi d\textbf{v}.
\end{equation}

Using conditions (\ref{svojstvo_O1}-\ref{svojstvo_O3}) we get,
\begin{equation}
	\label{dokaz2}
\int (\Psi -Y) \hat{O} (\Psi -Y)d\textbf{v}\geq0,
\end{equation}

\begin{equation}
	\label{dokaz3}
\int \Psi \hat{O} \Psi d\textbf{v}+\int Y \hat{O} Yd\textbf{v}- \int \Psi \hat{O} Yd\textbf{v}- \int Y \hat{O} \Psi d\textbf{v}=\int \Psi \hat{O} \Psi d\textbf{v}+\int Y \hat{O} Yd\textbf{v}-2\int Y \hat{O} \Psi d\textbf{v} \geq0,
\end{equation}
\small
\begin{equation}
	\label{dokaz4}
\int \Psi \hat{O} \Psi d\textbf{v}+\int Y \hat{O} Yd\textbf{v}+2 \int YZd\textbf{v}=\int \Psi \hat{O} \Psi d\textbf{v}+\int Y \hat{O} Yd\textbf{v}-2\int Y \hat{O} Yd\textbf{v}=\int \Psi \hat{O} \Psi d\textbf{v}-\int Y \hat{O} Yd\textbf{v} \geq0,
\end{equation}
\normalsize
\begin{equation}
	\label{dokaz5}
\int \Psi \hat{O} \Psi d\textbf{v} \geq \int Y \hat{O} Yd\textbf{v}.
\end{equation}

The left-hand side is the entropy production of the rarefied gas.   It comes  from Eq. (\ref{dokaz5})  that  solutions of the Boltzmann equation are in accordance with MEP principle.

There is another approach to this problem which has been proposed by 
Kohler \cite{kohlerM}.
  Due to the fact that Eq. (\ref{entropy}) is proportional to the  heat flux he 
 had  fixed the  heat flux and varied the distribution function under constraint 
\begin{equation}
\label{heat}
 q_x=constant.
\end{equation}
We note that due to the already fixed temperature field the entropy production is also fixed.
Kohler did not impose stationary conditions on the trial distribution function. In his approach the entropy generated by molecular collisions need not be equal to the entropy produced by fluxes and  gradients. 
If one does
not take care of this unconsistency the conclusion can be reached, by using the calculus of variation, that solutions
of the Boltzmann equation are those that generate minimum entropy production

\section{Conclusion}

The Onsager principle of the least dissipation of energy is valid for  processes close to the equilibrium state. 
In this paper the equivalence between the Onsager principle of the least dissipation of energy and the MEP principle is established.
Starting from the fact that fluxes  are the main phenomenological   characteristic of irreversible processes 
we have expanded the density of entropy production as the function of fluxes up to the second order.
We have found  that   the dissipation function introduced by Onsager  is 
 the entropy production in the space of the fluxes.
 Invoking the first law of thermodynamics 
the equivalency between the Onsager principle of the least dissipation energy and the MEP principle is established.
 
 It follows from the Boltzmann equation that entropy production is quantitatively  related to the collision integral.
 Collisions between molecules in nonequilibrium state produce entropy. 
 In this paper the solutions of the linearized Boltzmann equations are considered. 
    It is found that these solutions correspond to the extremum of entropy production. The nature of the extremum depends on the constraints. 
     Assuming  that the entropy produced by molecular collisions is equal to the entropy productions due to the heat conductions or/and viscosity we  find that the solutions of the Boltzmann equation satisfy the MEP principle.
     
     The principle of minimum entropy production is valid if  one grants  fixed fluxes. However, the starting assumption of the fixed thermodynamic forces and the additional assumption 
     of fixed fluxes leave no room for variation. Although both approaches are equivalent from 
     the mathematical point of view  one has to notice that the assumption of fixed fluxes is not consistent with the  starting assumption of the stationary process.  Namely, the additional request that entropy produced by molecular collisions  need  not be  equal to the entropy production  described by thermodynamic forces contradicts the starting assumption of the stationary process.

In short, revisiting the linear nonequilibrium thermodynamics and linearized Boltzmann equation shows that both approaches are in accordance with the MEP principle. 
Generally we can conclude  that the MEP principle is valid for  processes close to the  equilibrium state.

\section*{Acknowledgements}   
The present work was  supported by the bilateral research project of the Slovenia-Croatia Cooperation in Science and Technology, 2009-2010 and Croatian Ministry of Science grant No. 177-1770495-0476  to DJ.


\bibliographystyle{mdpi}
\makeatletter
\renewcommand\@biblabel[1]{#1. }
\makeatother

\end{document}